\documentclass[aps,prl,showpacs,twocolumn,preprintnumbers,superscriptaddress,nofootinbib,floatfix,10pt]{revtex4-1}
\usepackage{amsfonts,amssymb,stmaryrd,latexsym,amsmath,braket}
\usepackage{mathrsfs,dsfont}
\usepackage{graphicx,subfigure}
\usepackage{comment}
\usepackage{times}
\usepackage{slashed}
\usepackage{bm}
\usepackage{braket}
\usepackage{chapterbib}
\usepackage{color}
\usepackage{longtable}
\usepackage{hyperref}
\hypersetup{
    colorlinks=true,
    linkcolor=blue,
    citecolor=blue,
    urlcolor=blue
}
\usepackage{bibunits}

\newcommand{\beginsupplement}{%
        \setcounter{table}{0}
        \renewcommand{\thetable}{S\arabic{table}}%
        \setcounter{figure}{0}
        \renewcommand{\thefigure}{S\arabic{figure}}%
        \setcounter{equation}{0}
        \renewcommand{\theequation}{S\arabic{equation}}%
     }

\makeatletter
\let\saved@includegraphics\includegraphics
\AtBeginDocument{\let\includegraphics\saved@includegraphics}
\makeatother

\graphicspath{{figures/}}
\begin{document}

\begin{bibunit}[apsrev4-2]

\title{From bare two-nucleon interaction to nuclear matter and finite nuclei in a relativistic framework}

\author{Shihang Shen}
\affiliation{Peng Huanwu Collaborative Center for Research and Education, International Institute for Interdisciplinary and Frontiers, Beihang University, Beijing 100191, China}

\author{Jun-Xu Lu}
\email[Corresponding author: ] {ljxwohool@buaa.edu.cn}
\affiliation{School of Physics, Beihang University, Beijing 102206, China}

\author{Li-Sheng Geng}
\email[Corresponding author: ]{lisheng.geng@buaa.edu.cn}
\affiliation{School of Physics, Beihang University, Beijing 102206, China}
\affiliation{Sino-French Carbon Neutrality Research Center, \'Ecole Centrale de P\'ekin/School
 of General Engineering, Beihang University, Beijing 100191, China}
\affiliation{Peng Huanwu Collaborative Center for Research and Education, International Institute for Interdisciplinary and Frontiers, Beihang University, Beijing 100191, China}
\affiliation{Beijing Key Laboratory of Advanced Nuclear Materials and Physics, Beihang University, Beijing 102206, China}
\affiliation{Southern Center for Nuclear-Science Theory (SCNT), Institute of Modern Physics, Chinese Academy of Sciences, Huizhou 516000, China}

\author{Jie Meng}
\affiliation{State Key Laboratory of Nuclear Physics and Technology, School of Physics, Peking University, Beijing 100871, China}

\author{Wei-Jiang Zou}
\affiliation{State Key Laboratory of Nuclear Physics and Technology, School of Physics, Peking University, Beijing 100871, China}

\begin{abstract}
Understanding nuclear forces, infinite nuclear matter, and finite nuclei within a unified framework has remained a central challenge
in nuclear physics for decades. While most \textit{ab initio} studies employ nonrelativistic Schr\"odinger-equation frameworks,
this work offers a relativistic perspective. Using a leading-order (LO) relativistic chiral interaction, we describe two-nucleon scattering
via the Thompson equation, symmetric nuclear matter, and medium-mass nuclei (Ca, Ni, Zr, Sn) via the relativistic Brueckner-Hartree-Fock theory.
Systematic uncertainties from regulator cutoffs and interaction parameters are analyzed.
The empirical saturation region of nuclear matter is reproduced, and the binding energies and charge radii of medium-mass nuclei agree reasonably well with experimental data, significantly improving the ``Coester line".
These results highlight that the relativistic approach, employing a leading-order chiral force with only four low-energy constants and no three-nucleon forces, can capture the most important dynamics and offer a complementary pathway to address longstanding challenges in nuclear \textit{ab initio} studies.
\end{abstract}
\maketitle
\date{today}

\section{Introduction}

Understanding quantum systems from fundamental interactions and equations, i.e.,  \textit{ab initio},
is a foundational task across diverse fields, from hadron physics~\cite{Berges:2020fwq}, nuclear physics~\cite{Hergert:2020bxy},
condensed matter~\cite{yamazaki2025,holle2025}, to quantum chemistry~\cite{Ruggenthaler:2022ayz}.
In nuclear systems, due to the complexity of nuclear forces and strong correlations among a large number of nucleons, ranging from two to over two hundred,
\textit{ab initio} calculations are particularly challenging and remain a rapidly evolving field of study.

Starting from the basic symmetries of quantum chromodynamics, chiral effective field theory (EFT) provides a solid
foundation for modern nuclear forces~\cite{Weinberg:1990rz,Epelbaum:2008ga,Machleidt:2011zz,Epelbaum:2019kcf}.
Techniques such as renormalization group~\cite{Bogner:2003wn,Bogner:2006pc} and
wavefunction matching methods~\cite{Elhatisari:2022zrb} can help address numerical challenges arising from the strong repulsive core of nuclear interactions.
Numerous \textit{ab initio} frameworks have advanced the understanding of nuclear structure and reaction to unprecedented levels~\cite{Hammer:2019poc}, including
Brueckner-Goldstone theory \cite{Hu:2016nkw,Logoteta:2016nzc,Sammarruca:2014zia,Li:2012zzq,Day:1967zza},
coupled cluster method~\cite{Xu:2024lsu,Hu:2021trw,Novario:2021low,Morris:2017vxi,Hagen:2015yea,Hagen:2013nca},
Gamow shell model~\cite{Zhang:2025vct,Michel:2021jkx,Li:2019pmg},
in-medium similarity renormalization group (IMSRG)~\cite{Xu:2024lsu,Hu:2021trw,Stroberg:2019bch,Yao:2019rck,Gebrerufael:2016xih,Lapoux:2016exf,Hergert:2015awm},
many-body perturbation theory~\cite{Alp:2025wjn,Hu:2021trw,Drischler:2021kxf,Miyagi:2021pdc,Tichai:2020dna},
Monte-Carlo shell model~\cite{Otsuka:2022bcf,Shimizu:2012mv,Liu:2011xv,otsuka2001monte},
no-core shell model~\cite{Hebborn:2022iiz,Launey:2021sua,Hupin:2018biv,Gebrerufael:2016xih,Barrett:2013nh,Navratil:2000ww},
nuclear lattice EFT (NLEFT)~\cite{Lee:2025req,Shen:2024qzi,Elhatisari:2022zrb,Shen:2022bak,Lu:2021tab,Lahde:2019npb,Elhatisari:2015iga},
quantum Monte Carlo~\cite{Chambers-Wall:2024fha,Gnech:2023prs,Yang:2022rlw,Novario:2021low,Lynn:2019rdt,Carlson:2014vla},
self-consistent Green’s function method~\cite{Soma:2020xhv,Rios:2020oad,Arthuis:2020toz,Lapoux:2016exf}, and so on.

Despite remarkable successes in these nuclear \textit{ab initio} studies, challenges persist in achieving a unified description spanning
two-nucleon scattering, finite nuclei, and infinite nuclear matter~\cite{Elhatisari:2022zrb,Machleidt:2024bwl}.
In the nonrelativistic framework, three-nucleon forces (3NFs) are essential for reproducing the saturation of symmetric nuclear matter
(SNM) and the binding energies/sizes of medium mass nuclei.
See, for example, good descriptions achieved by recent NLEFT studies~\cite{Elhatisari:2022zrb,Niu:2025uxk}.
However, consistent and accurate treatment of 3NFs remains challenging~\cite{Sammarruca:2020jsp,Hebeler:2020ocj,Tews:2020hgp}.
Pioneering relativistic Brueckner-Hartree-Fock (RBHF) studies suggest that SNM saturation can be significantly
improved even with only the two-nucleon force (2NF)~\cite{Anastasio:1980jm,Brockmann:1990cn}, with connections between relativistic effects
and 3NFs in the nonrelativistic framework extensively explored~\cite{Brown:1985gt,Forest:1995sg,Sammarruca:2012vb,Muther:2016cod}.
Similar improvements in finite nuclei using phenomenological 2NFs have been reported in local density approximations ~\cite{Muther:1988cp,brockmann1992,Muther:2016cod},
self-consistent finite-basis studies for medium-mass nuclei~\cite{Shen:2016bva,Shen:2017vqr,Shen:2019dls},
and coordinate-space calculations for light nuclei~\cite{Yang:2024wsg}.
These advances have motivated the development of covariant chiral forces~\cite{Ren:2016jna,Xiao:2018jot,Lu:2021gsb},
which can reproduce the neutron-proton phase shifts up to 200 MeV at the next-to-next-to-leading order (N$^2$LO)~\cite{Lu:2021gsb}.
Notably, SNM saturation is achievable using leading-order (LO) relativistic chiral forces~\cite{Zou:2023quo,Zou:2025dzh,Zou:2025dao}.

With the successful establishment of the relativistic chiral force and the description of nuclear matter, it is timely to explore a unified description of finite nuclei within the relativistic framework.
In this work, we employ the LO relativistic chiral force with a local regulator~\cite{Zou:2025dzh} to study phase shifts, SNM,
and medium-mass nuclei ($A = 40$ to $120$). With only four low-energy constants (LECs) determined by two-nucleon scattering data,
we demonstrate that the relativistic framework offers a simple yet powerful and unified tool to capture essential dynamics across diverse nuclear systems at the 2NF level, providing new insights into the origins of nuclear binding ~\cite{Lu:2018bat,Gnech:2023prs}.

\section{Formalism}

The LO relativistic chiral force contains one-pion exchange and four contact terms
\begin{align}\label{eq:VLO}
  V =& \frac{g_A^2}{4f_\pi^2} \frac{\gamma^\mu\gamma_5q_\mu\gamma^\nu\gamma_5q_\nu}{q^\mu q_\mu - m_\pi^2}
    + C_{\rm S} \mathds{1}\mathds{1}
    + C_{\rm V} \gamma^\mu \gamma_\mu \notag \\
  & + C_{\rm AV} \gamma^\mu\gamma_5 \gamma_\mu\gamma_5
    + C_{\rm T} \sigma^{\mu\nu}\sigma_{\mu\nu},
\end{align}
with $f_\pi = 92.4$ MeV and $g_A = 1.29$ \cite{ParticleDataGroup:2016lqr}, $m_\pi$ the pion mass,
$q = (E_{p'}-E_{p}, \mathbf{p}'-\mathbf{p})$ the four momentum transfer, $\lambda$ the helicity
quantum number. 
Here we adopt the covariant power counting rule in which the expansion parameters are $s-4M^2=-(p_1-p_2)^2\sim\mathcal{O}(p^2)$, $t=(p_1-p_3)^2\sim \mathcal{O}(p^2)$, $m_\pi\sim\mathcal{O}(p)$ and $M\sim\mathcal{O}(p^0)$. See \cite{SM} for more details.
$C_{\rm S,V,AV,T}$ are the four LECs
to be determined by fitting to scattering phase shifts.
A local momentum-space regulator $\exp(-\mathbf{q}^2/\Lambda^2)$ with cutoff $\Lambda$ is
applied~\cite{Zou:2025dzh}. 

For many-body calculations, the relativistic Brueckner-Hartree-Fock theory~\cite{Brockmann:1990cn,Tong:2018qwx,Wang:2021mvg,Shen:2017vqr} is adopted.
The single-particle (s.p.) spinor in nuclear matter can be written as
\begin{equation}\label{eq:}
  \tilde{u}(\mathbf{p},s) = \left( \frac{\tilde{E}+\tilde{M}}{2\tilde{M}} \right)^{1/2}
  \left(\begin{array}{cc}
  1 \\ \frac{2\lambda p}{\tilde{E}+\tilde{M}}
  \end{array}\right)
  \chi_s,
\end{equation}
where the in-medium effective energy and mass are defined by
\begin{equation}\label{eq:}
  \tilde{M} = M + U_{\rm S}, \quad
  \tilde{E} = E - U_{\rm 0}.
\end{equation}
The scalar and zero-component vector potential, $U_{\rm S}$ and $U_{\rm 0}$,
are computed using the antisymmetrized $G$-matrix
\begin{equation}\label{eq:up}
  \hat{U}(\mathbf{p}) = \int_{|\mathbf{p}'|\leq p_F} d^3 \mathbf{p}'
  \frac{\tilde{M}_{\mathbf{p}}\tilde{M}_{\mathbf{p}'}}{\tilde{E}_{\mathbf{p}}\tilde{E}_{\mathbf{p}'}}
  \langle \mathbf{p}\mathbf{p}'|\bar{G}|\mathbf{p}\mathbf{p}' \rangle
  \approx \frac{\tilde{M}}{\tilde{E}} U_{\rm S} + U_{\rm 0}.
\end{equation}
In principle, the single-particle potential also includes space components and momentum dependence,
but these are neglected here due to their minor contributions~\cite{Brockmann:1990cn}.
The $G$-matrix is obtained by solving the in-medium Thompson equation
\begin{align}\label{eq:Gnm}
  G(\mathbf{q}',\mathbf{q}|\mathbf{P},W) =& V(\mathbf{q}',\mathbf{q})
  + \int \frac{d\mathbf{k}}{(2\pi)^3} V(\mathbf{q}',\mathbf{k})
  \frac{\tilde{M}^2}{\tilde{E}_{\mathbf{P}/2+\mathbf{k}}^2} \notag \\
  &\times \frac{Q(\mathbf{k},\mathbf{P})}{W-2\tilde{E}_{\mathbf{P}/2+\mathbf{k}}}
  G(\mathbf{k},\mathbf{q}|\mathbf{P},W),
\end{align}
where $W$ is the starting energy, and the Pauli operator
$Q$ restricts intermediate states to unoccupied levels.
A new $G$-matrix leads to updated s.p. potentials (or wave functions) in Eq.~(\ref{eq:up}), which will be
solved iteratively~\cite{Brockmann:1990cn}.
Recent improvements include the removal of the angle-averaging approximation in the calculation of energy~\cite{Tong:2018qwx},
and solving the RBHF equation in full Dirac space~\cite{Wang:2021mvg}.

For spherical finite nuclei, the s.p. wave function can be expressed as
\begin{equation}\label{eq:}
  \psi_a(\mathbf{r}) = \frac{1}{r}
  \left(\begin{array}{cc}
  F_{n_a\kappa_a}(r)\mathcal{Y}_{j_am_a}^{l_a}(\theta,\varphi) \\
  iG_{n_a\kappa_a}(r)\mathcal{Y}_{j_am_a}^{\tilde{l}_a}(\theta,\varphi)
  \end{array}\right),
\end{equation}
where $\mathcal{Y}$ denotes spinor spherical harmonics.
The quantum number $\kappa$ is defined as
$\kappa = \pm(j + 1/2)$ for $j = l \mp 1/2$, and $\tilde{l} = 2j - l$.
$F(r)$ and $G(r)$ are the radial wave functions.
The Bethe-Goldstone equation is solved in this basis as
\begin{align}\label{eq:Gfn}
  \langle a'b'|G|ab \rangle =& \langle a'b'|V|ab \rangle + \sum_{cd} \langle a'b'|V|cd \rangle \notag \\
  &\times \frac{Q(c,d)}{W-e_c-e_d} \langle cd|G|ab \rangle,
\end{align}
with the Pauli operator restricting intermediate states
$c$ and $d$ to unoccupied levels, and $e$ denotes the single-particle energy.
For details including the starting energy $W$, see Ref.~\cite{Shen:2017vqr}.
The Bethe-Goldstone equation is solved directly using bare forces from Eq.~(\ref{eq:VLO}),
and its convergence will be shown.
With the s.p. potential calculated from the $G$-matrix
\begin{equation}\label{eq:}
  \langle a|U|b \rangle = \sum_{c=1}^A \langle ac|\bar{G}|bc \rangle,
\end{equation}
a new set of s.p. wave functions is obtained by diagonalizing the s.p. Hamiltonian matrix.
More details are provided in the Supplemental Material~\cite{SM}.

\section{Results and discussion}

The four LECs in Eq.~(\ref{eq:VLO}) are determined by fitting the
$J \leq 1, np$-scattering phase shifts~\cite{Stoks:1993tb} up to 100 MeV~\cite{Zou:2025dzh}.
Cutoffs $\Lambda = 500, 600$, and $700$ MeV are considered. For each
cutoff, two parameter sets near the $\tilde{\chi}^2 = \sum_i^N \left[ \delta_i^{} - \delta_i^{(\rm PWA93)}\right]^2$ minimum
are selected to estimate systematic errors. Results are listed in Table~\ref{tab:para}.
See \cite{SM} for the discussion of the natural size criterion.
The phase shift descriptions improve with energy cutoffs, with \( \Lambda = 700 \, \text{MeV} \) performing the best.
We have verified that the deviation increases with further increases in the cutoff energy~\cite{Zou:2025dzh}.

\begin{table}[!htp]
    \centering
    \caption{LECs in units of GeV$^{-2}$ of different cutoffs.}
    \begin{tabular}{l|ccccc|c}
    \hline
    \hline
    Set & $\Lambda$ (MeV) & $C_{\rm S}$ & $C_{\rm V}$ & $C_{\rm AV}$ & $C_{\rm T}$ & $\tilde{\chi}^2/N$ \\
    \hline
    $\Lambda$500-I  & 500 & $-691.26$ & 603.38 & $-187.96$ & $-87.56$  & 7.6 \\
    $\Lambda$500-II & 500 & $-749.77$ & 664.46 & $-217.41$ & $-101.72$ & 7.4 \\
    $\Lambda$600-I  & 600 & $-672.28$ & 616.35 & $-194.20$ & $-87.52$  & 4.2 \\
    $\Lambda$600-II & 600 & $-711.64$ & 659.47 & $-216.41$ & $-98.15$  & 4.1 \\
    $\Lambda$700-I  & 700 & $-656.98$ & 614.94 & $-122.34$ & $-51.84$  & 3.6 \\
    $\Lambda$700-II & 700 & $-674.85$ & 637.83 & $-145.61$ & $-62.93$  & 3.3 \\
    \hline
    \hline
    \end{tabular}
    \label{tab:para}
\end{table}

\begin{figure}[!htbp]
  \includegraphics[width=0.5\textwidth]{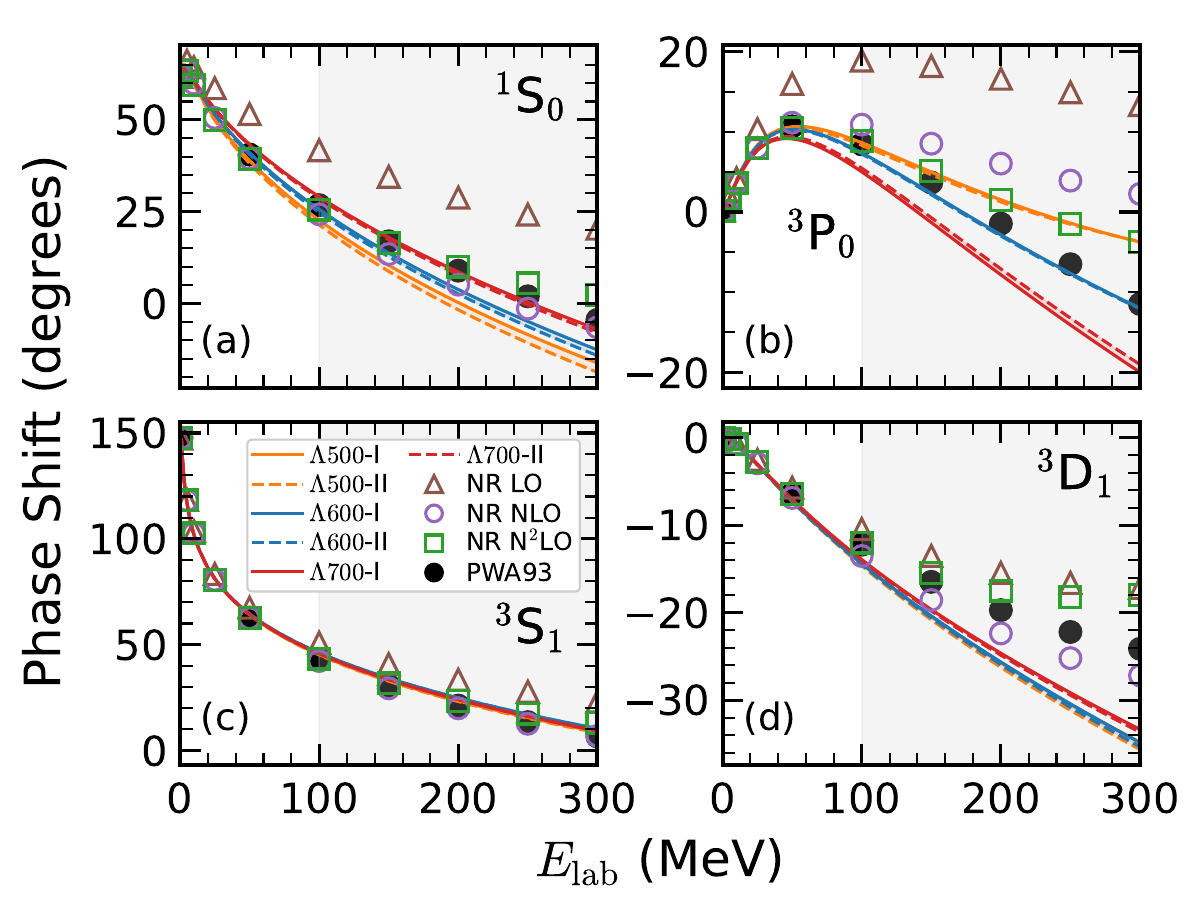}
  \caption{Selected $np$ phase shifts calculated using the LO relativistic chiral force (with cutoffs of 500–700 MeV),
  compared to the nonrelativistic (NR) chiral forces~\cite{Epelbaum:2014sza,Epelbaum:2014efa} and PWA93 data~\cite{Stoks:1993tb}.
  White backgrounds indicate regions used for LEC fitting, and gray regions are predictions.}
  \label{fig:phase}
\end{figure}

In Fig.~\ref{fig:phase}, selected partial waves calculated using the LO relativistic chiral force are
compared to the nonrelativistic results~\cite{Epelbaum:2014sza,Epelbaum:2014efa} and PWA93 data~\cite{Stoks:1993tb}.
Other $J \leq 4$ channels with comparisons to nonlocal regulators, as well as raw data in all figures of this work, can be found in the Supplemental Material~\cite{SM}.  
The LO relativistic chiral force achieves remarkable agreement with data, reaching up to 300 MeV in $S$-waves.  
A good agreement is also observed in the $^3\text{P}_0$ and $^3\text{D}_1$ channels up to 100 MeV.
We emphasize that this work does not aim for high-precision chiral interactions, but rather seeks to determine whether a unified description of nucleon-nucleon scattering, nuclear matter, and bulk properties of medium-mass nuclei can be achieved within a relativistic framework. In nonrelativistic frameworks, chiral forces have achieved remarkable precision at N$^4$LO \cite{Entem:2017gor, Reinert:2017usi}. Comparatively, the relativistic chiral framework currently only reaches N$^2$LO \cite{Lu:2021gsb}, with ongoing development toward higher orders.

\begin{figure}[!thbp]
  \includegraphics[width=0.4\textwidth]{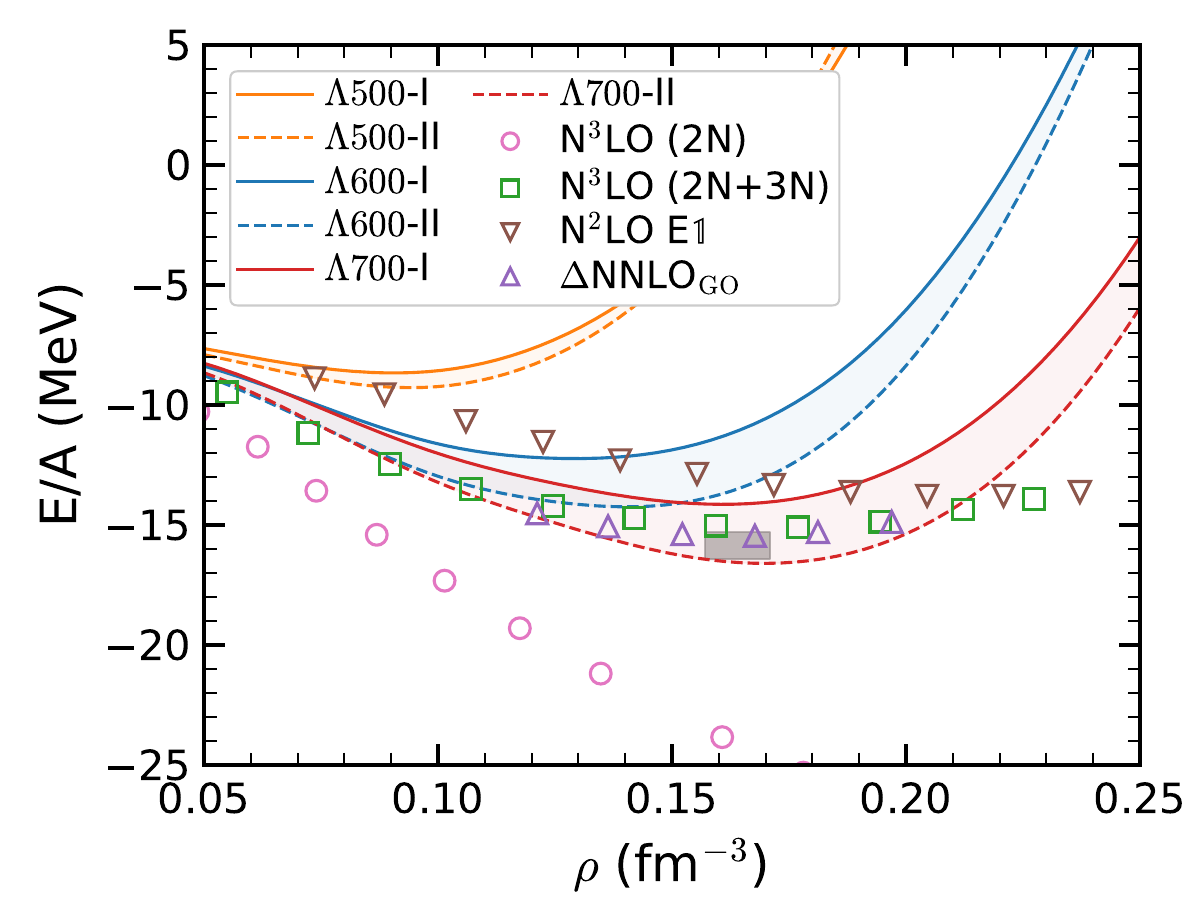}
  \caption{EoS of SNM from RBHF using the LO relativistic chiral force (with cutoffs of 500–700 MeV),
  compared to the nonrelativistic results~\cite{Sammarruca:2020jsp,Lonardoni:2019ypg,Jiang:2020the}.}
  \label{fig:eos}
\end{figure}

These forces are applied to symmetric nuclear matter using the RBHF theory~\cite{Zou:2025dzh},
yielding equations of state (EoS) for symmetric nuclear matter shown in Fig.~\ref{fig:eos}.
The empirical saturation region ($\rho = 0.164 \pm 0.007$ fm$^{-3}$, $E/A = -15.86 \pm 0.57$ MeV~\cite{Drischler:2017wtt}) is shaded gray.
The relativistic LO chiral force always yields SNM saturation, with lower cutoffs resulting in lower saturation densities and binding energies.
Note that the nonrelativistic studies require 3NFs for saturation (symbols in Fig.~\ref{fig:eos}~\cite{Sammarruca:2020jsp,Lonardoni:2019ypg,Jiang:2020the}),
while relativistic effects mimic key 3NF contributions~\cite{Brown:1985gt,Sammarruca:2012vb,Muther:2016cod}, reducing 3NF uncertainties.
For the study of pure neutron matter, see Ref.~\cite{Zou:2025dzh}.

Despite similar phase shifts for a given cutoff (Fig.~\ref{fig:phase} and Table~\ref{tab:para}),
EoS uncertainties persist.
This may arise from the fact that the interactions were determined solely by nucleon-nucleon (NN) scattering,
with antinucleon degrees of freedom unconstrained.
For the NN scattering observable, this has no effect, but for nuclear medium,
the nucleon-antinucleon excitation provides important relativistic effects
similar to a repulsive 3NF in the nonrelativistic framework that leads to 
saturation~\cite{Brown:1985gt,Sammarruca:2012vb,Muther:2016cod}.
Inclusion of antinucleons may help constrain such uncertainties and will be left for future
investigations.

Finite nuclei studies use the same LO relativistic chiral force (Table~\ref{tab:para}), focusing on medium-mass nuclei (\( ^{40}\text{Ca} \) to \( ^{120}\text{Sn} \)).
The Dirac Woods-Saxon basis~\cite{Zhou:2003jv} is solved in a spherical box $R = 7.5 \sim 8.8$ fm,
with s.p. angular momentum cutoff $l_{\rm cut} = 15$ to $22 \hbar$ and two-particle coupled angular momentum up to $J_{\rm cut} = 24$ for $^{120}$Sn. The largest size of two-body matrix in Eq.~(\ref{eq:Gfn}) for a given coupled angular momentum, parity, and isospin reaches a dimension of 50000.
The center-of-mass corrections are applied via projection before variation,
and the particle-state starting energies are set to $e' = e_{1s_{1/2}}^{(\nu)}$.
Numerical details and uncertainties can be found in Ref.~\cite{Shen:2017vqr}.

\begin{figure*}[!htbp]
  \includegraphics[width=0.24\textwidth]{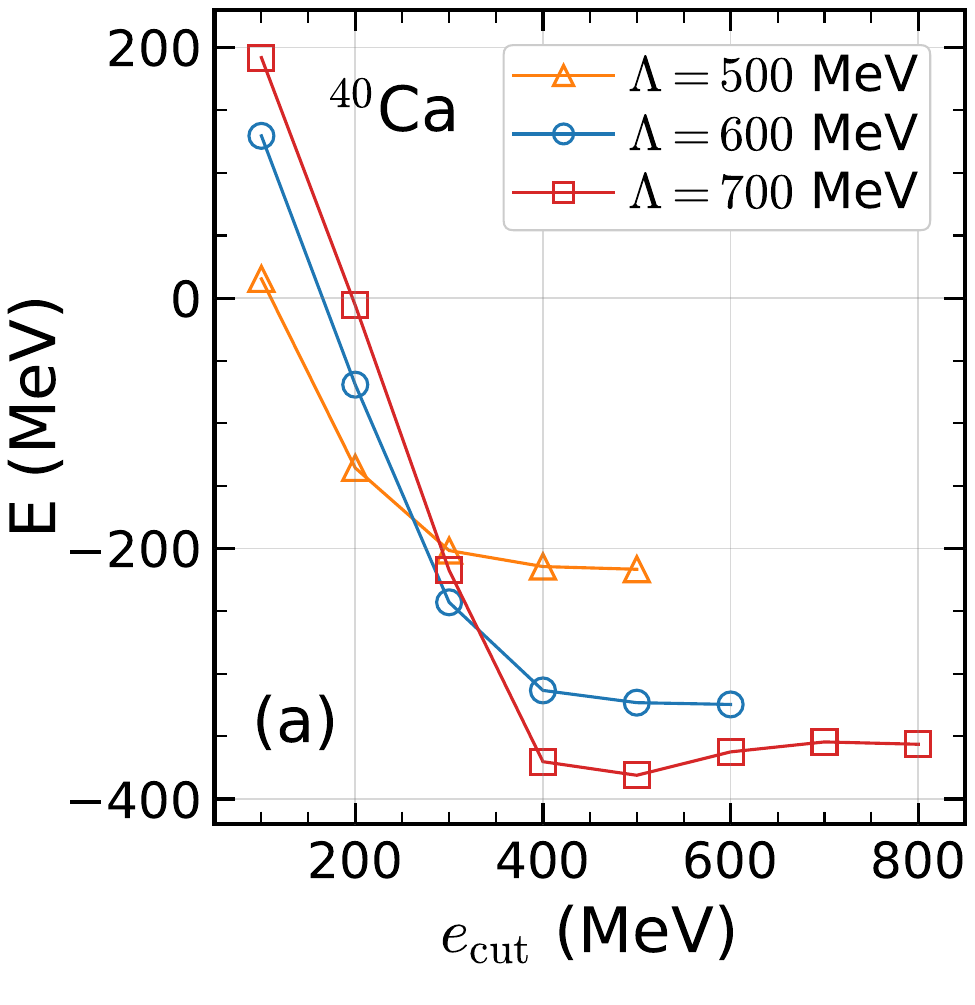}
  \includegraphics[width=0.75\textwidth]{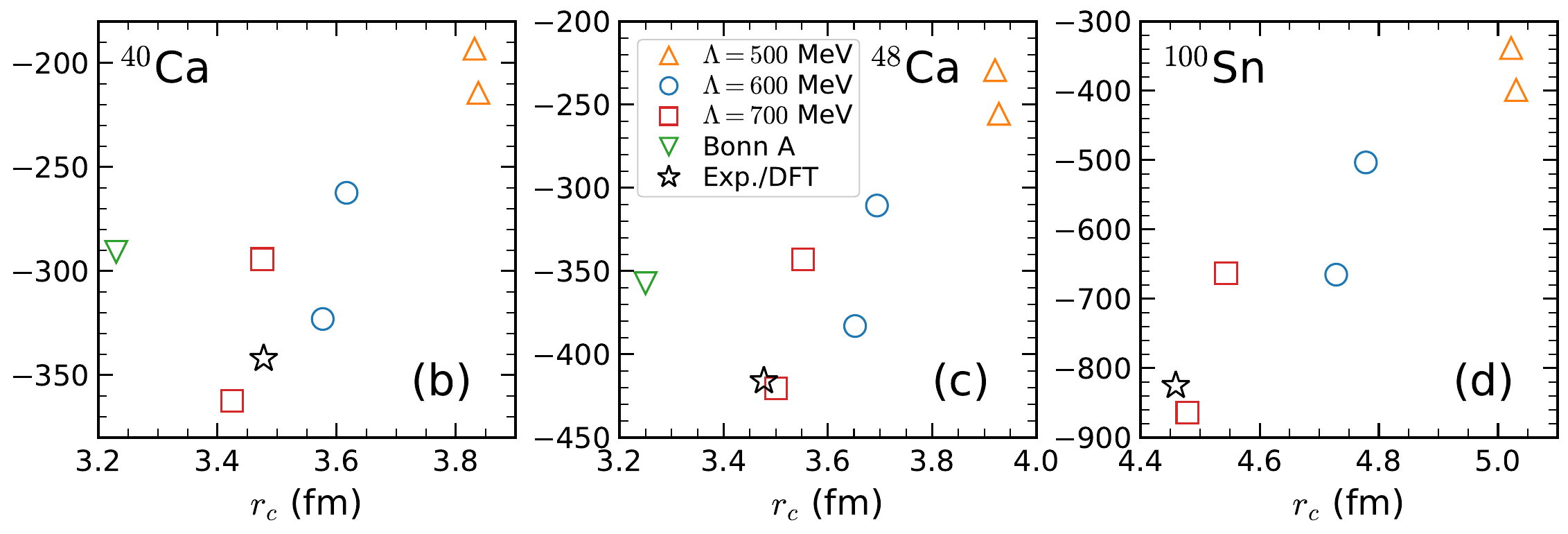}
  \caption{(a) Convergence of $^{40}$Ca energy (RBHF with the LO relativistic chiral force) vs. s.p. basis energy cutoff.
  (b-d) Energies and charge radii of $^{40}$Ca, $^{48}$Ca, and $^{100}$Sn, calculated using the LO relativistic chiral force (upper/lower symbols: sets I/II)
  vs. Bonn A~\cite{Shen:2017vqr,Shen:2018euq} and experimental data~\cite{Wang:2021xhn,Angeli:2013epw}.
  For the $^{100}$Sn radius, the theoretical prediction from density functional theory (DFT) studies~\cite{DRHBcMassTable:2024nvk} is used.}
  \label{fig:chk}
\end{figure*}

Fig.~\ref{fig:chk}(a) shows convergence for $^{40}$Ca (set II).
With s.p. energy cutoff $e_{\rm cut} < 100$ MeV, the nuclei are unbound, indicating the strong repulsive core in the bare forces.
The convergence is achieved near the expected cutoff energy $\Lambda = 500 \sim 700$ MeV.
Smaller cutoff values correspond to softer interactions.
Compared to large cutoff potentials like Bonn~\cite{Machleidt:1989tm} (convergence $\gtrsim 1$ GeV~\cite{Shen:2016bva,Shen:2017vqr}),
the chiral forces prioritize low-energy observables, offering computational efficiency.
This enables us to extend our study to heavier nuclei such as the Tin isotopes.

Fig.~\ref{fig:chk}(b-d) displays energies and charge radii for $^{40}$Ca, $^{48}$Ca, and $^{100}$Sn.
The LO relativistic results form bands encompassing experimental data, similar to SNM (Fig.~\ref{fig:eos}).
Compared to the ``Coester line" obtained in nonrelativistic BHF~\cite{Coester:1970ai},
RBHF with Bonn potentials improves the description of SNM~\cite{Brockmann:1990cn} and finite nuclei~\cite{Shen:2017vqr},
and the relativistic chiral forces further align with the data.

Two caveats remain, though:
1. RBHF (or the nonrelativistic BHF) framework represents the lowest-order of the Brueckner-Goldsteon hole-line expansion~\cite{Day:1967zza},
and includes only two-body correlations.
Many-body corrections require further rigorous studies, see, for instance,
three-body corrections in IMSRG~\cite{Heinz:2021xir},
three-particle three-hole corrections in coupled cluster method~\cite{Hagen:2016uwj}.
2. While the LO relativistic chiral force captures key physics in scattering, SNM, and nuclei, higher-order effects need to be explored for
unified improvements and uncertainty quantification for chiral expansions. See \cite{SM} for estimation and more discussion of uncertainties.

\begin{figure}[!htbp]
  \includegraphics[width=0.5\textwidth]{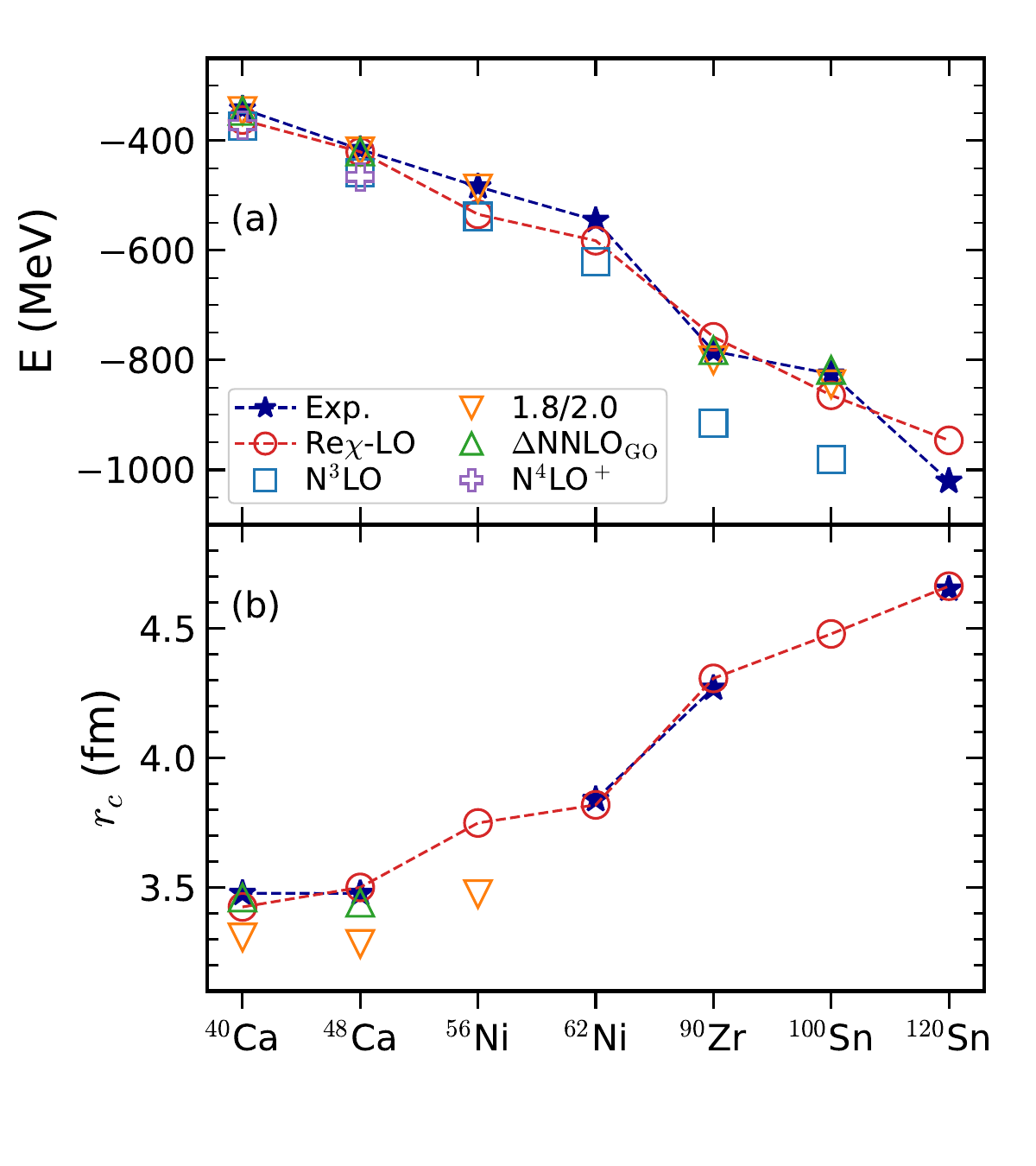}
  \caption{Energies (a) and charge radii (b) of nuclei ($A = 40$ to $120$) from RBHF using the LO relativistic chiral force (Re$\chi$-LO, $\Lambda$700-II),
  compared to the nonrelativistic chiral forces~\cite{Binder:2013xaa,Morris:2017vxi,Simonis:2017dny,Jiang:2020the,LENPIC:2022cyu}
  and experimental data~\cite{Wang:2021xhn,Angeli:2013epw}.}
  \label{fig:er700}
\end{figure}

With the LO relativistic chiral force offering a promising starting point,
we believe challenging questions remain in relativistic \textit{ab initio} studies.
By choosing the $\Lambda$700-II parameter set,
which provides the best overall description across nuclear systems, we investigate selected nuclei from
$^{40}$Ca to $^{120}$Sn.
The resulting energies and charge radii are shown in Fig.~\ref{fig:er700}, compared to the nonrelativistic
chiral forces~\cite{Binder:2013xaa,Morris:2017vxi,Simonis:2017dny,Jiang:2020the,LENPIC:2022cyu}
and experimental data~\cite{Wang:2021xhn,Angeli:2013epw}.
The overall agreement with the experimental data is remarkable, particularly when considering both energies and charge radii simultaneously.
Medium-mass nuclei have been extensively studied in the nonrelativistic ab initio
frameworks \cite{Binder:2013xaa,Ekstrom:2015rta,Morris:2017vxi,Simonis:2017dny,Hoppe:2019uyw,Huther:2019ont,Ekstrom:2017koy,Hergert:2012nb,Heinz:2024juw,Elhatisari:2022zrb},
yet achieving precise simultaneous reproduction of energies and radii remains challenging.
Systematic uncertainties, particularly those tied to 3NFs, remain significant.
By employing purely 2NFs at leading order, the relativistic framework can replicate key bulk nuclear properties,
showing the importance of relativistic effects in nuclear systems.

\section{Summary and discussion}

We have investigated two-nucleon scattering, symmetric nuclear matter, and finite nuclei using the 
Relativistic Brueckner-Hartree-Fock theory with the leading-order relativistic chiral force.
Systematic uncertainties arising from cutoff variations and parameter sets at fixed cutoffs are quantified.
Phase shifts, saturation properties, energies, and charge radii are described with reasonable accuracy,
demonstrating that the relativistic framework captures important physics and
establishes a robust foundation for further studies.
This framework can help in understanding the three-nucleon forces.
Future efforts could focus on two directions: advancing relativistic \textit{ab initio} many-body frameworks
and employing higher-order relativistic chiral forces.
Additionally, truncation uncertainties can be studied~\cite{Furnstahl:2015rha}, and 
low-momentum forces can be implemented to offer computational advantages for heavy nuclei~\cite{Wang:2023zdc,Huang:2025two}.


\section{Acknowledgments}
We gratefully acknowledge insightful discussions with Ulf-G. Mei\ss ner.
We also thank Evgeny Epelbaum for providing the nonrelativistic chiral force phase shift code.
This work is supported by the National Natural Science Foundation of China under Grant No. 12435007.
JM acknowledges support from the National Science Foundation of China under Grant No. 12435006
and the National Key R\&D Program of China under Grant No. NST202401016.

\putbib[bref-rbhf]

\end{bibunit}

\clearpage

\beginsupplement
\begin{bibunit}[apsrev4-2]
%
%




\begin{onecolumngrid}
\renewcommand{\thefigure}{S\arabic{figure}}
\renewcommand{\thetable}{S\arabic{table}}
\renewcommand{\theequation}{S\arabic{equation}}
\setcounter{figure}{0}
\setcounter{equation}{0}
\setcounter{table}{0}

\section*{Supplementary Material}

\subsection{Formalism}

\subsubsection{Power counting}

In this section, we briefly introduce our covariant power counting rule. In principle, the power counting rule is quite similar to that in the covariant $\pi N$ amplitude~\cite{Chen:2012nx}. 

In the covariant $\pi N$ interaction, the most general amplitudes reads
\begin{equation}
\begin{split}
\mathcal{T}&=\bar{u}_f \left(A^\pm+\frac{\slashed{q} + \slashed{q}'}{2}B^\pm\right)u_i  \\
           &=\bar{u}_f(D^\pm+\frac{i}{2M}\sigma^{\mu\nu}q_\mu'q_\nu B^\pm)u_i
\end{split}
\end{equation}
where the $u_f, u_i$ the spinors of incoming and outgoing nucleons, $q,q'$ the 4-momentum of incoming and outgoing pions, and $A^\pm, B^\pm, D^pm$ are the scalar functions composed of Mandelstam symbols $s,t,u$.

For such a covariant $\pi N$ amplitudes, the expansion parameters are chosen to be 
\begin{equation}
   s-M^2\sim\mathcal{O}(p),~t\sim\mathcal{O}(p^2),~m_\pi\sim\mathcal{O}(p), ~M\sim\mathcal{O}(p^0), \nu=\frac{s-u}{4M}\sim\mathcal{O}(p)
\end{equation}
and the chiral order of Lorentz scalar is counted as 
\begin{equation}
   \bar{u}_fu_i\sim\mathcal{O}(p^0), ~~\bar{u}_f \sigma^{\mu\nu} q_\mu q'_\nu u_i\sim \mathcal{O}(p^2)
\end{equation}

Quite similarly, the covariant potential for nucleon-nucleon interaction could be expressed as the product of scalar integral and various combination of bilinear similar to $\pi N$ amplitude but more complicated, which reads
\begin{equation}
     \mathcal{A}=L(s,t,M,m_\pi) \otimes B(p_1,p_2,p_3,p_4) 
\end{equation}
where $L$ the scalar integral composed of Mandelstam symbols and $B(p_1,p_2,p_3,p_4)$ are the different combinations of bilinears composed the 4-momentum of incoming and outgoing nucleons.

For the scalar integral parts, the expansion parameters are chosen as
\begin{equation}
   s-4M^2=-(p_1-p_2)^2\sim\mathcal{O}(p^2), ~t=(p_1-p_3)^2\sim \mathcal{O}(p^2), ~ m_\pi\sim\mathcal{O}(p),~ M\sim\mathcal{O}(p^0)
\end{equation}
And for the Bilinear part, one needs to treat each terms separately, such as
\begin{equation}
   \bar{u}_3u_1\bar{u}_4u_2\sim\mathcal{O}(p^0), \bar{u}_3(\slashed{p}_1-M)u_1\bar{u}_4u_2\sim\mathcal{O}(p^1)
\end{equation}

The power counting rules holds for loop diagrams since the variables in the amplitudes are again the Mandelstam symbols besides the masses of pions and nucleons. For an explicit formula for TPE we refer to Eq.~(7), Eq.~(8) in ref.~\cite{Xiao:2020ozd}.

Such a covariant power counting is consistent with Weinberg's power counting rules in the sense that the covariant rules could be reduced to the non-relativistic ones after the expansion of $1/M$ with $M$ the nucleon mass. But the difference is that the covariant blocks contain not only the corresponding non-relativistic terms at the same order, but also more terms due to relativistic energy-momentum relation, which are nominally of higher order. But the covariant power counting rules reorganize the interactions and ``promote" these terms to lower order in a self-consistent manner.

\subsubsection{Free space}

Rewrite the single-particle (s.p.) spinor in a simplified notation
\begin{equation}\label{eq:sm-u1}
  u(\mathbf{p},\lambda) = \sqrt{\frac{W}{2M}}
\left(\begin{array}{cc}
  1 \\ \frac{2\lambda |\mathbf{p}|}{W}
\end{array}\right) |\lambda\rangle,\quad \text{with}~W = E + M, \quad E^2 = \mathbf{p}^2 + M^2.
\end{equation}
Consider the center-of-mass frame, let the initial two interacting nucleons have momentum
$(\mathbf{p},-\mathbf{p})$ and final ones $(\mathbf{p}',-\mathbf{p}')$.
Together with the helicity quantum number, the interaction matrix element will be denoted as
\begin{equation}\label{eq:}
  \langle \lambda_1'\lambda_2'|V(\mathbf{p}',\mathbf{p}) | \lambda_1\lambda_2 \rangle.
\end{equation}
The isospin spinor will be omited for simplicity when not necessary.

The matrix element for one-pion-exchange (OPE) term is
\begin{align}\label{eq:sm-vope}
  & \langle \lambda_1'\lambda_2'|V_{\rm OPE}(\mathbf{p}',\mathbf{p}) | \lambda_1\lambda_2 \rangle \notag \\
  =& \frac{g_A^2}{4f_\pi^2} \frac{\left[ \bar{u}(\mathbf{p}',\lambda_1',t_1')\bm{\tau}\gamma^\mu\gamma_5
  q_\mu u(\mathbf{p},\lambda_1,t_1)\right] \cdot \left[ \bar{u}(-\mathbf{p}',\lambda_2',t_2')\bm{\tau}\gamma^\nu\gamma_5
  q_\nu u(-\mathbf{p},\lambda_2,t_2)\right]}{(E_{p'}-E_p)^2-(\mathbf{p}'-\mathbf{p})^2-m_\pi^2} \\
  =& \frac{g_A^2}{4f_\pi^2} \frac{W'}{2M} \frac{W}{2M} \left[ \left( \frac{2\lambda_1'p'}{W'} + \frac{2\lambda_1p}{W}
\right) (E_{p'}-E_p)
  - \left( 1 + \frac{4\lambda_1'\lambda_1^{}p'p}{W'W} \right) (2\lambda_1'p'-2\lambda_1p) \right] \notag \\
  & \times \left[ \left( \frac{2\lambda_2'p'}{W'} + \frac{2\lambda_2p}{W} \right) (E_{p'}-E_p)
  + \left( 1 + \frac{4\lambda_2'\lambda_2^{}p'p}{W'W} \right) (2\lambda_2'p'-2\lambda_2p) \right]
  \frac{\langle \lambda_1'\lambda_2'|\lambda_1^{}\lambda_2^{} \rangle
  \langle t_1't_2'|\bm{\tau}_1^{}\cdot \bm{\tau}_2^{} |t_1t_2\rangle}{
  (E_{p'}-E_p)^2-(\mathbf{p}'-\mathbf{p})^2-m_\pi^2}.
\end{align}
The retardation effect is ignored in current study therefore $E_{p'}-E_{p} \to 0$.
The matrix element for scalar term is
\begin{align}\label{eq:}
  \langle \lambda_1'\lambda_2'|V_{\rm S}(\mathbf{p}',\mathbf{p}) | \lambda_1\lambda_2 \rangle =&
  C_{\rm S} \left[ \bar{u}(\mathbf{p}',\lambda_1') \mathds{1} u(\mathbf{p},\lambda_1)\right]
  \left[ \bar{u}(-\mathbf{p}',\lambda_2') \mathds{1} u(-\mathbf{p},\lambda_2)\right] \notag \\
  =& C_{\rm S} \frac{W'}{2M} \frac{W}{2M} \left( 1 - \frac{4\lambda_1'\lambda_1^{}p'p}{W'W} \right)
  \left( 1 - \frac{4\lambda_2'\lambda_2^{}p'p}{W'W} \right) \langle \lambda_1'\lambda_2'|\lambda_1^{}\lambda_2^{}
\rangle.
\end{align}
The matrix element for vector term is
\begin{align}\label{eq:}
  \langle \lambda_1'\lambda_2'|V_{\rm V}(\mathbf{p}',\mathbf{p}) | \lambda_1\lambda_2 \rangle =&
  C_{\rm V} \left[ \bar{u}(\mathbf{p}',\lambda_1') \gamma^\mu u(\mathbf{p},\lambda_1)\right]
  \left[ \bar{u}(-\mathbf{p}',\lambda_2') \gamma_\mu u(-\mathbf{p},\lambda_2)\right] \notag \\
  =& C_{\rm V} \frac{W'}{2M} \frac{W}{2M} \left( 1 + \frac{4\lambda_1'\lambda_1^{}p'p}{W'W} \right)
  \left( 1 + \frac{4\lambda_2'\lambda_2^{}p'p}{W'W} \right) \langle \lambda_1'\lambda_2'|\lambda_1^{}\lambda_2^{}
\rangle \notag \\
  &- C_{\rm V} \frac{W'}{2M} \frac{W}{2M} \left( \frac{2\lambda_1'p'}{W'} + \frac{2\lambda_1p}{W} \right)
  \left( \frac{2\lambda_2'p'}{W'} + \frac{2\lambda_2p}{W} \right) \langle \lambda_1'\lambda_2'|
\bm{\sigma}_1\cdot\bm{\sigma}_2 | \lambda_1^{}\lambda_2^{} \rangle.
\end{align}
The matrix element for axial vector term is
\begin{align}\label{eq:}
  \langle \lambda_1'\lambda_2'|V_{\rm AV}(\mathbf{p}',\mathbf{p}) | \lambda_1\lambda_2 \rangle =&
  C_{\rm AV} \left[ \bar{u}(\mathbf{p}',\lambda_1') \gamma^\mu \gamma_5 u(\mathbf{p},\lambda_1)\right]
  \left[ \bar{u}(-\mathbf{p}',\lambda_2') \gamma_\mu \gamma_5 u(-\mathbf{p},\lambda_2)\right] \notag \\
  =& C_{\rm AV} \frac{W'}{2M} \frac{W}{2M} \left( \frac{2\lambda_1'p'}{W'} + \frac{2\lambda_1p}{W} \right)
  \left( \frac{2\lambda_2'p'}{W'} + \frac{2\lambda_2p}{W} \right) \langle
\lambda_1'\lambda_2'|\lambda_1^{}\lambda_2^{} \rangle \notag \\
  &- C_{\rm AV} \frac{W'}{2M} \frac{W}{2M} \left( 1 + \frac{4\lambda_1'\lambda_1^{}p'p}{W'W} \right)
  \left( 1 + \frac{4\lambda_2'\lambda_2^{}p'p}{W'W} \right) \langle \lambda_1'\lambda_2'|
\bm{\sigma}_1\cdot\bm{\sigma}_2 | \lambda_1^{}\lambda_2^{} \rangle.
\end{align}
The matrix element for tensor term is
\begin{align}\label{eq:sm-vt}
  \langle \lambda_1'\lambda_2'|V_{\rm T}(\mathbf{p}',\mathbf{p}) | \lambda_1\lambda_2 \rangle =&
  C_{\rm T} \left[ \bar{u}(\mathbf{p}',\lambda_1') \sigma^{\mu\nu} u(\mathbf{p},\lambda_1)\right]
  \left[ \bar{u}(-\mathbf{p}',\lambda_2') \sigma_{\mu\nu} u(-\mathbf{p},\lambda_2)\right] \notag \\
  =& 2C_{\rm T} \frac{W'}{2M} \frac{W}{2M} \left(-\frac{2\lambda_1'p'}{W'} + \frac{2\lambda_1p}{W} \right)
  \left(-\frac{2\lambda_2'p'}{W'} + \frac{2\lambda_2p}{W} \right) \langle \lambda_1'\lambda_2'|
  \bm{\sigma}_1\cdot\bm{\sigma}_2 |\lambda_1^{}\lambda_2^{} \rangle \notag \\
  &+ 2C_{\rm T} \frac{W'}{2M} \frac{W}{2M} \left( 1 - \frac{4\lambda_1'\lambda_1^{}p'p}{W'W} \right)
  \left( 1 - \frac{4\lambda_2'\lambda_2^{}p'p}{W'W} \right) \langle \lambda_1'\lambda_2'|
  \bm{\sigma}_1\cdot\bm{\sigma}_2 | \lambda_1^{}\lambda_2^{} \rangle.
\end{align}

In this work, a local form of momentum regulator is multiplied to the two-body matrix element
\begin{equation}\label{eq:sm-fq}
  f_{\rm local}(\mathbf{p}',\mathbf{p}) = e^{-(\mathbf{p}'-\mathbf{p})^2/\Lambda^2},
\end{equation}
where $\Lambda$ is the cutoff.

For the rest of solution of Thmposon equation in helicity basis and
calculation of partial wave phase shift, the reader is refered to
Refs.~\cite{Erkelenz:1971caz,Machleidt:1987hj}

\subsubsection{Symmetric nuclear matter}

In symmetric nuclear matter (SNM), the s.p. spinor is replaced by
\begin{equation}\label{eq:sm-u2}
  u(\mathbf{p},\lambda) = \sqrt{\frac{\tilde{W}}{2\tilde{M}}}
\left(\begin{array}{cc}
  1 \\ \frac{2\lambda |\mathbf{p}|}{\tilde{W}}
\end{array}\right) |\lambda\rangle,\quad \text{with}~\tilde{W} = \tilde{E} + \tilde{M}, \quad \tilde{E}^2 = \mathbf{p}^2 + \tilde{M}^2.
\end{equation}
where the effective energy and mass are defined as
\begin{equation}\label{eq:sm-tilde}
  \tilde{M} = M + U_{\rm S}, \quad \tilde{E} = E - U_{\rm 0},
\end{equation}
with s.p. potential
\begin{equation}\label{eq:}
  \hat{U}(\mathbf{p}) = \int_{|\mathbf{p}'|\leq p_F} d^3 \mathbf{p}'
  \frac{\tilde{M}_{\mathbf{p}}\tilde{M}_{\mathbf{p}'}}{\tilde{E}_{\mathbf{p}}\tilde{E}_{\mathbf{p}'}}
  \langle \mathbf{p}\mathbf{p}'|\bar{G}|\mathbf{p}\mathbf{p}' \rangle
  \approx \frac{\tilde{M}}{\tilde{E}} U_{\rm S} + U_{\rm 0}.
\end{equation}
In principle the single-particle potential also has space component and momentum dependence,
but due to the small contribution \cite{Brockmann:1990cn} they are ingored here.
The two-body interaction matrix elements in SNM are similar as in free space in
Eqs.~(\ref{eq:sm-vope}-\ref{eq:sm-vt}), with replacing the energy and mass to the one
with s.p. potentials (\ref{eq:sm-u2},\ref{eq:sm-tilde}).

For the rest of solution of $G$-matrix and calculation of s.p. potential,
the reader is refered to Refs.~\cite{Brockmann:1990cn}

\subsubsection{Spherical nuclei}

For finite nuclei with spherical symmetry, the s.p. wave function can be written as
\begin{equation}\label{eq:}
  \psi_a(\mathbf{r}) = \frac{1}{r}
  \left(\begin{array}{cc}
  F_{n_a\kappa_a}(r)\mathcal{Y}_{j_am_a}^{l_a}(\theta,\varphi) \\
  iG_{n_a\kappa_a}(r)\mathcal{Y}_{j_am_a}^{\tilde{l}_a}(\theta,\varphi)
  \end{array}\right),
\end{equation}
The two-body interaction matrix element is
\begin{equation}\label{eq:sm-vabcd}
  \langle ab|V|cd \rangle = \int \frac{d\mathbf{q}}{(2\pi)^3} f_{\rm local}(\mathbf{q})
  \int d\mathbf{r}_1 d\mathbf{r}_2 e^{i\mathbf{q}\cdot(\mathbf{r}_1-\mathbf{r}_2)}
  \bar{\psi}_a(\mathbf{r}_1) \bar{\psi}_b(\mathbf{r}_2) \Gamma(1,2)
  {\psi}_c(\mathbf{r}_1) {\psi}_d(\mathbf{r}_2),
\end{equation}
with $f_{\rm local}$ defined in Eq.~(\ref{eq:sm-fq}), and $\Gamma(1,2)$ is the Lorentz operator
for different terms.
For OPE term there is a propagator which is also in the local form of momentum
transfer $\mathbf{q}$.
While a nonlocal form regulator
\begin{equation}\label{eq:}
  f_{\rm nonlocal}(\mathbf{p}',\mathbf{p}) = e^{-({\mathbf{p}'}^2+\mathbf{p}^2)/\Lambda^2}
\end{equation}
makes little difference for the calculation of two-body interaction matrix element
in free space or SNM (\ref{eq:sm-vope}-\ref{eq:sm-vt}),
it is less straightforward to be applied in the finite nuclei (\ref{eq:sm-vabcd})
and a local form is much simpler.

Define the particle-hole $jj$-coupled matrix element
\begin{equation}\label{eq:}
  \langle ab|V|cd \rangle^J = \sum_{m_am_bm_cm_d} (-1)^{j_c-m_c}(-1)^{j_b-m_b}
  C_{j_am_aj_c-m_c}^{JM} C_{j_dm_dj_b-m_b}^{JM} \langle ab|V|cd \rangle.
\end{equation}
The matrix element for OPE term is
\begin{align}\label{eq:}
  \langle ab|V_{\rm OPE}|cd \rangle^J =& -\frac{g_A^2}{4f_\pi^2} \frac{2}{\pi} \frac{(-1)^{j_b-j_d}}{(2J+1)^2}
  \int_0^\infty \frac{q^4 dq}{m_\pi^2+q^2} e^{-q^2/\Lambda^2} \notag \\
  &\times \langle a||\sqrt{J+1} j_{J+1}[Y_{J+1}\sigma]_J + \sqrt{J} j_{J-1} [Y_{J-1}\sigma]_J ||c \rangle
          \langle b||\sqrt{J+1} j_{J+1}[Y_{J+1}\sigma]_J + \sqrt{J} j_{J-1} [Y_{J-1}\sigma]_J ||d \rangle,
\end{align}
with
\begin{equation}\label{eq:}
  \langle a||j_L[Y_L\sigma]_J||c \rangle = \langle j_al_a||[Y_L\sigma]_J||j_cl_c \rangle
  \int dr F_a^*(r) j_L(qr) F_c(r) + \langle j_a\tilde{l}_a||[Y_L\sigma]_J||j_c\tilde{l}_c \rangle
  \int dr G_a^*(r) j_L(qr) G_c(r),
\end{equation}
and $j_L(qr)$ is the spherical Bessel function.

The matrix element for scalar term is
\begin{align}\label{eq:}
  \langle ab|V_{\rm S}|cd \rangle^J =& C_{\rm S} \frac{2}{\pi} \frac{(-1)^{j_b-j_d}}{2J+1}
  \int_0^\infty q^2 dq e^{-q^2/\Lambda^2} \langle a||\gamma^0j_{J}Y_{J} ||c \rangle \langle b||\gamma^0j_{J}Y_{J} ||d \rangle,
\end{align}
with
\begin{equation}\label{eq:}
  \langle a||\gamma^0j_JY_J||c \rangle = \langle j_al_a||Y_J||j_cl_c \rangle
  \int dr F_a^*(r) j_J(qr) F_c(r) + \langle j_a\tilde{l}_a||Y_J||j_c\tilde{l}_c \rangle
  \int dr G_a^*(r) j_J(qr) G_c(r).
\end{equation}

The matrix element for vector term is
\begin{align}\label{eq:}
  \langle ab|V_{\rm V}|cd \rangle^J =& C_{\rm V} \frac{2}{\pi} \frac{1}{2J+1}
  \int_0^\infty q^2 dq e^{-q^2/\Lambda^2} \left[ (-1)^{j_b-j_d} I_{\rm V0}(q) - (-1)^{j_b+j_d+J} I_{\rm V1}(q)\right],
\end{align}
with
\begin{subequations}\label{eq:}\begin{align}
I_{\rm V0}(q) =& \left[
\langle a||Y_J||c \rangle \int dr F_a^*(r) j_J(qr) F_c(r) +
\langle \tilde{a}||Y_J||\tilde{c} \rangle \int dr G_a^*(r) j_J(qr) G_c(r) \right] \notag \\
& \times \left[
\langle b||Y_J||d \rangle \int dr F_b^*(r) j_J(qr) F_d(r) +
\langle \tilde{b}||Y_J||\tilde{d} \rangle \int dr G_b^*(r) j_J(qr) G_d(r) \right], \\
I_{\rm V1}(q) =& -\sum_L (-1)^L \left[-
\langle \tilde{a}|| [Y_L\sigma]_J ||c \rangle \int dr G_a^*(r) j_L(qr) F_c(r) +
\langle {a}|| [Y_L\sigma]_J ||\tilde{c} \rangle \int dr F_a^*(r) j_L(qr) G_c(r) \right] \notag \\
& \times \left[-
\langle \tilde{b}|| [Y_L\sigma]_J ||d \rangle \int dr G_b^*(r) j_L(qr) F_d(r) +
\langle {b}|| [Y_L\sigma]_J ||\tilde{d} \rangle \int dr F_b^*(r) j_L(qr) G_d(r) \right].
\end{align}\end{subequations}

The matrix element for axial vector term is
\begin{align}\label{eq:}
  \langle ab|V_{\rm AV}|cd \rangle^J =& C_{\rm AV} \frac{2}{\pi} \frac{1}{2J+1}
  \int_0^\infty q^2 dq e^{-q^2/\Lambda^2} \left[ (-1)^{j_b-j_d} I_{\rm AV0}(q) - (-1)^{j_b+j_d+J} I_{\rm AV1}(q)\right],
\end{align}
with
\begin{subequations}\label{eq:}\begin{align}
I_{\rm AV0}(q) =& -\left[-
\langle \tilde{a}||Y_J||c \rangle \int dr G_a^*(r) j_J(qr) F_c(r) +
\langle {a}||Y_J||\tilde{c} \rangle \int dr F_a^*(r) j_J(qr) G_c(r) \right] \notag \\
& \times \left[-
\langle \tilde{b}||Y_J||d \rangle \int dr G_b^*(r) j_J(qr) F_d(r) +
\langle {b}||Y_J||\tilde{d} \rangle \int dr F_b^*(r) j_J(qr) G_d(r) \right], \\
I_{\rm AV1}(q) =& \sum_L (-1)^L \left[
\langle {a}|| [Y_L\sigma]_J ||c \rangle \int dr F_a^*(r) j_L(qr) F_c(r) +
\langle \tilde{a}|| [Y_L\sigma]_J ||\tilde{c} \rangle \int dr G_a^*(r) j_L(qr) G_c(r) \right] \notag \\
& \times \left[
\langle {b}|| [Y_L\sigma]_J ||d \rangle \int dr F_b^*(r) j_L(qr) F_d(r) +
\langle \tilde{b}|| [Y_L\sigma]_J ||\tilde{d} \rangle \int dr G_b^*(r) j_L(qr) G_d(r) \right].
\end{align}\end{subequations}

The matrix element for tensor term is
\begin{align}\label{eq:}
  \langle ab|V_{\rm T}|cd \rangle^J =& C_{\rm T} \frac{2}{\pi} \frac{1}{2J+1}
  \int_0^\infty q^2 dq e^{-q^2/\Lambda^2} \left[ (-1)^{j_b-j_d} 2 I_{\rm T0}(q) + (-1)^{j_b+j_d+J} 2 I_{\rm T1}(q)\right],
\end{align}
with
\begin{subequations}\label{eq:}\begin{align}
I_{\rm T0}(q) =& -\sum_L (-1)^L \left[
\langle \tilde{a}|| [Y_L\sigma]_J ||c \rangle \int dr G_a^*(r) j_L(qr) F_c(r) +
\langle {a}|| [Y_L\sigma]_J ||\tilde{c} \rangle \int dr F_a^*(r) j_L(qr) G_c(r) \right] \notag \\
& \times \left[
\langle \tilde{b}|| [Y_L\sigma]_J ||d \rangle \int dr G_b^*(r) j_L(qr) F_d(r) +
\langle {b}|| [Y_L\sigma]_J ||\tilde{d} \rangle \int dr F_b^*(r) j_L(qr) G_d(r) \right], \\
I_{\rm T1}(q) =& \sum_L (-1)^L \left[
\langle {a}|| [Y_L\sigma]_J ||c \rangle \int dr F_a^*(r) j_L(qr) F_c(r) -
\langle \tilde{a}|| [Y_L\sigma]_J ||\tilde{c} \rangle \int dr G_a^*(r) j_L(qr) G_c(r) \right] \notag \\
& \times \left[
\langle {b}|| [Y_L\sigma]_J ||d \rangle \int dr F_b^*(r) j_L(qr) F_d(r) -
\langle \tilde{b}|| [Y_L\sigma]_J ||\tilde{d} \rangle \int dr G_b^*(r) j_L(qr) G_d(r) \right].
\end{align}\end{subequations}

For the rest of solution of $G$-matrix and RHF equation,
the reader is refered to Refs.~\cite{Shen:2017vqr}.

\subsection{Phase shift}

In Fig.~\ref{fig:sm-phase} we show the $np$ scattering phase shifts
for partial waves with $J \leq 4$ calculated
using relativistic local LO chiral forces from cutoff 500 to 700 MeV.
The results are compared with relativistic nonlocal LO chiral forces \cite{Zou:2023quo},
Bonn forces \cite{Machleidt:1989tm}, nonrelativistic N$^4$LO+ \cite{Reinert:2017usi}, and PWA93 data \cite{Stoks:1993tb}.
Overall the descriptiton of relativistic LO chiral forces using local or nonlocal regulator
gives similar results.
For some partial waves the local form gives better description such as $^{3}$D$_1$ and
$\varepsilon_1$, while for $^{1}$P$_1$ the nonlocal form is better.
The cutoff dependence using nonlocal regulator is generally weaker than using local regulator
in current study.

\begin{figure}[!htbp]
\centering
\includegraphics[width=1.0\textwidth]{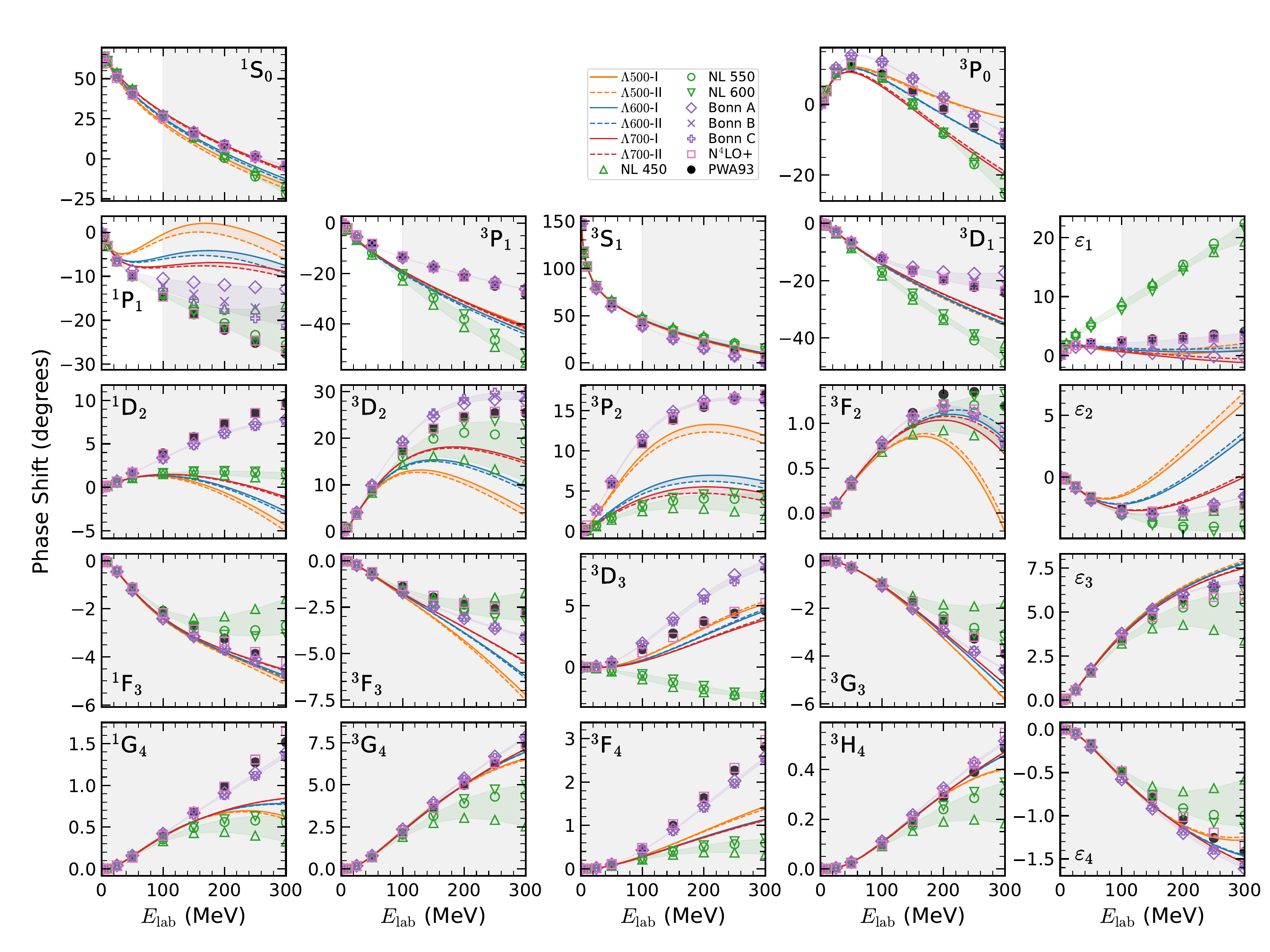}
\caption{$np$ phase shifts for partial waves with $J \leq 2$ calculated
using relativistic local LO chiral forces from cutoff 500 to 700 MeV, in comparison with
relativistic nonlocal (NL) LO chiral forces from cutoff 450 to 600 MeV \cite{Zou:2023quo},
Bonn forces \cite{Machleidt:1989tm}, nonrelativistic N$^4$LO+ \cite{Reinert:2017usi}, and PWA93 data \cite{Stoks:1993tb}.
White background is region where data are used to fit the LECs in current work and gray region
are predictions.}
\label{fig:sm-phase}
\end{figure}

\subsection{Natural size condition of LECs}

The natural size for LECs is closely related to the power counting rules adopted to organize the interaction. 
In the present situation, the covariant power counting rules dictate that in addition to the non-relativistic operators of the corresponding order, it also contains operators of higher orders as per the non-relativistic power counting rules. 
Therefore, the $C_{\rm S,V,AV,T}$ here are a combination of contributions from various orders of hard scale $\Lambda$, as shown in the Table IV of \cite{Xiao:2018jot}. 
Consequently, the estimation of their natural size is non-trivial, and we simply take the non-relativistic estimation in Ref.~\cite{Epelbaum:2001fm,Epelbaum:2014efa} as rough guidelines.

The natural size of LECs depends on power counting rules. 
 The Lagrangian for a system of pions and nucleons interacting through derivatives ($\partial^\mu$) acting on any field
 can be written in the schematic form
 \begin{align}
 \mathcal{L} &= \mathcal{L}_{\mathrm{free}} + \Delta \mathcal{L}\\
 \mathcal{L}_{\mathrm{free}}&=\bar{\psi}(i\gamma_\mu\partial^\mu-M)\psi+\frac{1}{2}(\partial^\mu\bm{\pi})^2-\frac{1}{2}m_{\bm{\pi}}^2\bm{\pi}^2\\
 \Delta\mathcal{L}&\sim a\left(\frac{\bar{\psi}(\cdots)\psi}{b}\right)^l\left(\frac{\bm{\pi}}{c}\right)^m\left(\frac{\partial^\mu}{d}\right)^n.
 \end{align}
 A Lagrangian (basically an energy density) has dimension 4 (i.e., behaves as $[E^4]$). The combination $\bar{\psi}\psi$ has dimension $[E^3]$, while $\bm{\pi}$ and $\partial^\mu$ (as well as $m_\pi$ and $M$) have dimension $[E]$, and $\Delta\mathcal{L}$ of course has dimension $[E^4]$.

Our first assumption is that only the scales $f_\pi$ and $\Lambda$ occur in $a,b,c,d$ and also that $\Delta \mathcal{L}$ should reproduce $\mathcal{L}_{\mathrm{free}}$.
Clearly, $c$ must be $f_\pi$, since that quantity sets the scale for the pion field.
Setting $l=0$ and $m=n=2$ gives $a/c^2d^2\sim1$, while $m=0,l=1$, and $n=0,1$ give $a/b\sim \Lambda$ and $a/bd\sim1$ (where we have substituted $M\sim\Lambda$). 
These three equations can be uniquely solved to produce $d\sim\Lambda,
a\sim f_\pi^2\Lambda^2$, and $b\sim f_\pi^2\Lambda$ and thus~\cite{Friar:1996zw}
\begin{align}
\Delta \mathcal{L} = c_{lmn}\left(\frac{\bar{\psi}(\cdots)\psi}{f_\pi^2\Lambda}\right)^l\left(\frac{\bm{\pi}}{f_\pi}\right)^m\left(\frac{\partial^\mu}{\Lambda}\right)^n f_\pi^2\Lambda^2.
\end{align}
where $c_{lmn}$ is dimensionless. All Dirac matrices, nucleon isospin operators, etc. have been ignored and are indicated by the dots. 
The second assumption is that a reasonable theory should have $c_{lmn}\sim1$, or that the theory is "natural". This is also called naive dimensional power counting.

Taking the Lagrangian corresponding to $C_{\rm S}$ as an example, the estimation for $C_{\rm S}$ is as follows:
\begin{align}
C_{\rm S}(\bar{\psi}\psi)(\bar{\psi}\psi)\sim c_{200}f_\pi^2\Lambda^2
\left(\frac{\bar{\psi}\psi}{f_\pi^2\Lambda}\right)^2\Rightarrow
C_{\rm S}\sim \frac{1}{f_\pi^2}
\end{align}
Similarly, the magnitude estimates for the other remaining LECs can be obtained as follows: $C_{\rm V,AV,T}\sim\frac{1}{f_\pi^2}$. 
All currently LECs obtained satisfy the natural size condition.


\subsection{Uncertainty quantification}

Truncation errors in effective field theory can be quantified by the Bayesian model \cite{Furnstahl:2015rha}.
In this approach, an observable $X$ is written in terms of dimensionless expansion coefficients $c_i$ as
\begin{equation}
    X = X_{\rm ref} (c_0 + c_2Q^2 + c_3 Q^3 + \dots ).
\end{equation}
The chiral expansion parameter $Q$ is defined as
\begin{equation}
    Q = \rm{Max} \left\{ \frac{p}{\Lambda}, \frac{m_{\pi}}{\Lambda} \right\},
\end{equation}
where $p$ is the nucleon momentum and $\Lambda$ the momentum cutoff.
At NNLO, the reference value $X_{\rm ref}$ is defined as \cite{Epelbaum:2019zqc}:
\begin{equation}
    X_{\rm ref} = \rm{Max} \left\{ |X_{\rm LO}|, \frac{|X_{\rm LO}-X_{\rm NLO}|}{Q^2}, \frac{|X_{\rm NLO}-X_{\rm NNLO}|}{Q^2} \right\}.
\end{equation}
For current LO study, we use $X_{\rm ref} = |X_{\rm LO}|$.
With the assumption that the next chiral order dominates the truncation errors, the dimensionless residue at order $k$ reads
\begin{equation}
    \Delta_k = \sum_{n = k+1}^{\infty} c_n Q^n.
\end{equation}
The truncation uncertainty can be written as
\begin{equation}
    \Delta X = X_{\rm ref} \Delta_k.
\end{equation}
To estimate the magnitude of $\Delta X$, we assume $c_n = 1$, thus
\begin{equation}
    \Delta X = |X_{\rm LO}| (Q^2 + Q^3 + \dots ) = |X_{\rm LO}| \frac{Q^2}{1-Q} = |X_{\rm LO}| \frac{p^2}{\Lambda(\Lambda - p)}.
\end{equation}
Near the saturation density $\rho \approx 0.17$ fm$^{-3}$, the corresponding Fermi momentum is about 1.35 fm$^{-1}$ or 266 MeV.
Using momentum cutoff $\Lambda = 600$ MeV which lies in the middle of current study range, the uncertainty is estimated as
\begin{equation}
     \Delta X \sim |X_{\rm LO}| \times 35\%.
\end{equation}
For $^{100}$Sn presented in the Fig.~3 of main text, the energy using $\Lambda$-II set is $-665$ MeV and the uncertainty
$\Delta E$ will be about 233 MeV, around half of the energy difference 465 MeV from cutoff $\Lambda = 500$ to 700 MeV.
Results of symmetric nuclear matter calculation at NLO can reduce such uncertainty to nearly by half and shows promising convergence behavior \cite{Zou:2025dao}.

When higher order such as NLO calculated, more rigorous estimation can be made based on the Bayesian model \cite{Furnstahl:2015rha}.
With the assumption that the next $h$ chiral orders dominate the truncation errors, the the dimensionless residue
\begin{equation}
    \Delta_k \simeq \sum_{n=k+1}^{k+h} c_n Q^n
\end{equation}
obeys the following probability distribution
\begin{equation}
    P_h(\Delta|c_{i\leq k}) = \frac{\int_0^\infty d\bar{c} P_h(\Delta|\bar{c}) P(\bar{c}) \prod_{i\in A} P(c_i|\bar{c})}{\int_0^\infty d\bar{c} P(\bar{c}) \prod_{i\in A} P(c_i|\bar{c})},
\end{equation}
with $A = \left\{ n\in \mathbf{N} | n \leq k \wedge n \neq 1 \wedge n \neq m \right\}$, and
\begin{equation}
    P_h(\Delta|\bar{c}) = \delta\left( \Delta - \sum_{j = k+1}^{k+h} c_j Q^j \right) \prod_{i=k+1}^{k+h} \int_{-\infty}^{\infty} dc_i P(c_i|\bar{c}).
\end{equation}
$\bar{c}$ is the hyperparameter for the probability distribution function $P(c_i|\bar{c})$ which follows the log-uniform probability distribution as
\begin{equation}
    P(\bar{c}) = \frac{1}{\ln (\bar{c}_{>}/\bar{c}_{<})} \frac{1}{\bar{c}} \theta(\bar{c}-\bar{c}_{<}) \theta(\bar{c}_{>}-\bar{c}).
\end{equation}
Following \cite{Epelbaum:2019zqc}, a Gaussian prior is used
\begin{equation}
    P(c_i|\bar{c}) = \frac{1}{\sqrt{2\pi}\bar{c}} e^{-c_i^2/(2\bar{c}^2)},
\end{equation}
and the following values are adopted: $h = 10, \bar{c}_{<} = 0.5$, and $\bar{c}_{>}= 10.0$.
For any given degree-of-belief interval, the truncation uncertainties $\Delta X = X_{\rm ref} \Delta_k$ can be numerically obtained by integrating over $\Delta$.

\subsection{Numerical values}

The numerical values of energy and charge radius
calculated by RBHF using relativistic LO chiral force (set $\Lambda$700-II), shown
in Fig.~4 of the main text, are listed in
Table~\ref{tab:sm-er}, in comparison with experimental data \cite{Wang:2021xhn,Angeli:2013epw}.
All other numerical values of phase shift, equation of state, energy, and charge radius plotted in this work will be attached as raw data.

\begin{table}[!htp]
    \centering
    \caption{Energies (in MeV) and charge radii (in fm) of selected nuclei calculated by RBHF using relativistic
    LO chiral force (set $\Lambda$700-II), in comparison with experimental data \cite{Wang:2021xhn,Angeli:2013epw}.}
    \begin{tabular}{l|cc|cc}
    \hline
    \hline
    & \multicolumn{2}{c}{RBHF} & \multicolumn{2}{c}{Exp.} \\
    Nucleus & $E$ & $r_c$ & $E$ & $r_c$ \\
    \hline
    $^{40}$Ca  & $-362.3$ & 3.42 & $-342.1$  & 3.48  \\
    $^{48}$Ca  & $-420.4$ & 3.50 & $-416.0$  & 3.48  \\
    $^{56}$Ni  & $-534.3$ & 3.75 & $-484.0$  & --  \\
    $^{62}$Ni  & $-582.4$ & 3.82 & $-545.3$  & 3.84  \\
    $^{90}$Zr  & $-757.7$ & 4.31 & $-783.9$  & 4.27  \\
    $^{100}$Sn & $-864.2$ & 4.48 & $-825.2$  & --  \\
    $^{120}$Sn & $-946.3$ & 4.66 & $-1020.5$ & 4.65  \\
    \hline
    \hline
    \end{tabular}
    \label{tab:sm-er}
\end{table}

\end{onecolumngrid}

\putbib[bref-rbhf]

\end{bibunit}

\end{document}